\pdfoutput=1

\documentclass[num-refs]{nbdt-article}

\usepackage{siunitx}

\papertype{Original Article}
\paperfield{Analysis}

\title{Characterizing the nonlinear structure of shared variability in cortical neuron populations using latent variable models}

\author[1\authfn{1}]{Matthew R Whiteway}
\author[2,3,4\authfn{2}]{Karolina Socha}
\author[2,3,4]{Vincent Bonin}
\author[1,5]{Daniel A Butts}

\affil[1]{Program in Applied Mathematics \& Statistics, and Scientific Computation, University of Maryland, College Park, MD, United States}
\affil[2]{Neuro-Electronics Research Flanders, Leuven, Belgium}
\affil[3]{Department of Biology \& Leuven Brain Institute, KU Leuven, Leuven, Belgium}
\affil[4]{VIB, Leuven, Belgium}
\affil[5]{Department of Biology and Program in Neuroscience and Cognitive Sciences, University of Maryland, College Park, MD, United States}

\corraddress{Matthew R Whiteway}
\corremail{m.whiteway@columbia.edu}

\presentadd[\authfn{1}]{Zuckerman Institute, Columbia University, New York, NY, United States}
\presentadd[\authfn{2}]{University College London, London, UK}

\fundinginfo{This work was supported by the NSF IIS-1350990 and NIH R21 EY025403-02 (MRW and DAB) and NERF, a joint research initiative of imec, KU Leuven and VIB (KS and VB)}

\runningauthor{Whiteway et al.}

\begin{document}

\maketitle

\begin{abstract}
Sensory neurons often have variable responses to repeated presentations of the same stimulus, which can significantly degrade the stimulus information contained in those responses. This information can in principle be preserved if variability is shared across many neurons, but depends on the structure of the shared variability and its relationship to sensory encoding at the population level. The structure of this shared variability in neural activity can be characterized by latent variable models, although they have thus far typically been used under restrictive mathematical assumptions, such as assuming linear transformations between the latent variables and neural activity. Here we introduce two nonlinear latent variable models for analyzing large-scale neural recordings. We first present a general nonlinear latent variable model that is agnostic to the stimulus tuning properties of the individual neurons, and is hence well suited for exploring neural populations whose tuning properties are not well characterized. This motivates a second class of model, the Generalized Affine Model, which simultaneously determines each neuron’s stimulus selectivity and a set of latent variables that modulate these stimulus-driven responses both additively and multiplicatively. While these approaches can detect very general nonlinear relationships in shared neural variability, we find that neural activity recorded in anesthetized primary visual cortex (V1) is best described by a single additive and single multiplicative latent variable, i.e., an ``affine model''. In contrast, application of the same models to recordings in awake macaque prefrontal cortex discover more general nonlinearities to compactly describe the population response variability. These results thus demonstrate how nonlinear latent variable models can be used to describe population variability, and suggest that a range of methods is necessary to study different brain regions under different experimental conditions.

\end{abstract}


\section{Introduction}
The activity of sensory cortical neurons is highly variable in response to repeated presentations of the same stimulus \cite{tolhurst1983statistical, vogels1989response}. This single-neuron response variability has the potential to limit the information about the stimulus, since multiple stimuli could in principle elicit the same number of spikes. Nevertheless, at the population level this single-neuron variability would have minimal impact on sensory coding if it were due to noisy biological processes independent to each neuron \cite{faisal2008noise}, since averaging responses across a population would result in an unbiased estimate of the stimulus \cite{zohary1994correlated}. However, early work demonstrated that variability was instead shared among pairs of neurons (noise correlations; \cite{zohary1994correlated}), motivating both experimental \cite{cohen2011measuring} and theoretical \cite{kohn2016correlations} research into understanding the structure and implications of this shared variability for population coding.

Recently, data from a wide variety of recording modalities has demonstrated that a large portion of this variability in cortex is not only shared among pairs of neurons, but among much larger populations as well \cite{arieli1996dynamics, goris2014partitioning, ecker2014state, scholvinck2015cortical, lin2015nature, pachitariu2015state, rabinowitz2015attention, okun2015diverse, arandia2016multiplicative}. The shared variability in these large cortical populations has been successfully modeled using latent variable methods, where a small number of factors, or latent variables, drive neural activity across the entire population \cite{cunningham2014dimensionality}. Many popular latent variable methods such as Principal Component Analysis and Factor Analysis assume a linear mathematical form, where the activity of each latent variable contributes to neural responses independently of the other latent variables. These methods, in addition to reproducing the population activity, can account for salient features of noise correlations \cite{okun2015diverse}. However, linear methods cannot account for higher-order correlations, which may play an important role in population coding in cortex \cite{ohiorhenuan2010sparse, ohiorhenuan2011information, yu2011higher, koster2014modeling}. It is therefore necessary to characterize this higher-order statistical structure of shared variability with nonlinear latent variable models, which serves as a first step to understanding the impact of this structure on population coding. 

A large fraction of the recent work analyzing the structure of shared variability in large neural populations (both linear and nonlinear) has been applied to activity recorded from primary visual cortex (V1), where sensory-driven neural responses to drifting gratings are relatively well understood. In this context, shared variability in V1 responses has been modeled using a single additive latent variable that modulates neural activity across the entire population \cite{ecker2014state, okun2015diverse}, and also modeled using a multiplicative, rather than additive, latent variable that acts as a gain on the stimulus-driven response \cite{goris2014partitioning}. More recently, the ``affine'' model of Lin et al. (2015) – and the related multi-gain model of Arandia-Romero et al. (2016) – combined both additive and multiplicative latent variables to account for neural variability, which described more of the population response than either latent variable alone. However, fitting multiplicative latent variables is non-trivial, and thus far methods have employed restrictive assumptions in order to fit these models to data. As a result, it is currently unclear whether any of the proposed models are the best model of population response variability in V1, or simply the best under these assumptions. For example, it is possible that another form of nonlinear interaction between the stimulus-driven response and latent variables would be more appropriate than a multiplicative gain, or that multiple gain terms should be included \cite{rabinowitz2015attention}. 

To address these questions, we introduce a general nonlinear latent variable modeling framework that infers latent variables by transforming the high-dimensional neural activity into a low-dimensional set of latent variables using neural networks. In the simplest case, this transformation can be a linear projection, and when a separate linear projection maps the latent variable activity to the predicted neural activity, the resulting model is closely related to PCA and FA \cite{whiteway2017revealing}. Here, we consider nonlinear transformations, resulting in a general nonlinear latent variable model that can, in principle, model arbitrary nonlinear interactions between the latent variables. Furthermore, this framework can be adapted to include the affine model without any restrictive assumptions, and similarly extended to create a generalized version of the affine model with an arbitrary number of additive and multiplicative latent variables. By fitting these two classes of nonlinear latent variable models to population recordings from anesthetized monkey V1, we find that the population response variability is well described by the affine model, where one multiplicative and one additive latent variable can generally capture most of the variance of the more general models. We also fit these models to population recordings from awake macaque prefrontal cortex (PFC), and find that, unlike anesthetized V1, the affine model is not sufficient to capture most of the shared variability. Thus, this novel approach to fitting nonlinear latent variable models provides a new means to compare constrained latent variable models to more general models of population response variability, and takes a step towards elucidating the nonlinear computations that underlie neural activity.


\section{Results}

\begin{figure}[t]
  \vspace{20pt}
  \centering
  \centerline{\includegraphics[scale=0.85]{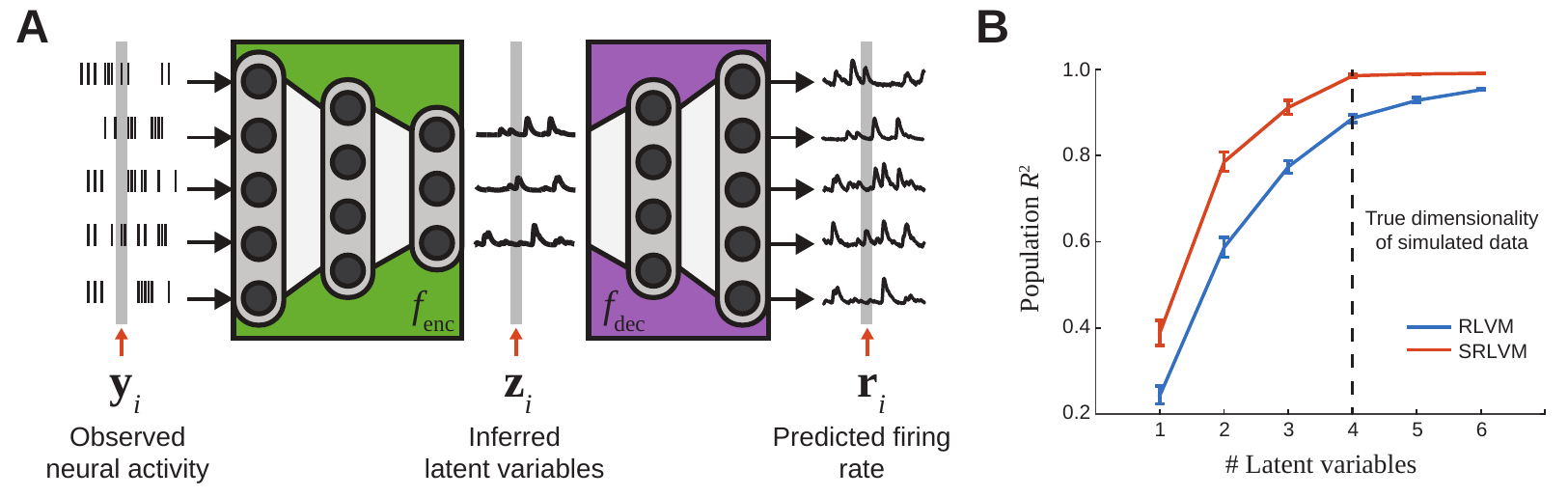}}
  \vspace{5pt}
  \caption{\linespread{1}\selectfont{} \textbf{The structure of the Stacked Rectified Latent Variable Model (SRLVM).} \textbf{A}: The SRLVM is a neural network that predicts the population activity $\mathbf{r}_i$ using the measured population activity $\mathbf{y}_i$. The high-dimensional measured population activity is transformed into a lower-dimensional set of latent variables $\mathbf{z}_i$ using an encoding function $f_{\text{enc}}$, which is learned from the data with a neural network (\emph{green box}). A decoding function $f_{\text{dec}}$ is simultaneously learned with a separate neural network (\emph{purple box}) to transform the latent variables into the predicted population activity. The RLVM models $f_{\text{enc}}$ and $f_{\text{dec}}$ as single-layer neural networks, whereas the SRLVM models $f_{\text{enc}}$ and $f_{\text{dec}}$ as multi-layer neural networks, allowing for more complex, nonlinear transformations of the data. \textbf{B}: Fits of the RLVM and SRLVM to simulated data generated using a nonlinear combination of four latent variables. The SRLVM is able to capture most of the variance using four latent variables, whereas the RLVM needs more latent variables to capture the effect of the nonlinear interactions. Error bars are standard error of the mean over ten cross-validation folds. \label{fig:srlvm}}
\end{figure}

\subsection{A general framework for fitting nonlinear latent variable models}

Latent variable models aim to describe the high-dimensional activity of many neurons using a low-dimensional set of latent variables. A linear latent variable model predicts the activity of each neuron as a linear combination of the latent variables, so that the vector of predicted neural responses $\mathbf{r}_i$ on trial $i$ is given by
\begin{equation} \label{eqn:linear-lv}
\mathbf{r}_i = W \mathbf{z}_i + \mathbf{b}
\end{equation}
where $\mathbf{z}_i$ is the vector of latent variables, $W$ is a matrix mapping the latent variables to the predicted responses, and $\mathbf{b}$ is a vector of biases. Because neural activity is stochastic, the predicted $\mathbf{r}_i$ represents the expected value of the activity on trial $i$; depending on the modeling scenario, $\mathbf{r}_i$ could be the mean parameter of a Gaussian distribution (e.g. spiking activity in large time bins, or fluorescence values from two-photon imaging) or the rate parameter of a Poisson distribution (e.g. spiking activity in small time bins).

Linear latent variable models such as Principal Component Analysis (PCA) and Factor Analysis (FA) – and variations thereof (e.g. dPCA \cite{kobak2016demixed}) – are finding widespread use in neuroscience due to the ease of fitting these models to neural data, their interpretability, and their flexibility \cite{cunningham2014dimensionality}. However, given the number of nonlinear cellular and circuit mechanisms governing neural variability (e.g. the discrete nature of the spike generation process), there is no reason to believe that relationships between latent variables and the activity of different neurons will be entirely linear. This issue is often circumvented in part by using linear latent variable models to analyze neural data only after activity has been averaged over repeated trials. Thus, while linear approaches have provided insights about the nature of neural computations during particular tasks (e.g. \cite{churchland2012neural, ahrens2012brain}), a deeper understanding of how neural activity is structured during single trials, for example to study neural activity during decision-making tasks or during learning, will require nonlinear latent variable methods.

A nonlinear latent variable model predicts the activity of each neuron as some nonlinear transformation $f_{\text{dec}}$ of the latent variables ($f_{\text{dec}}$ ``decodes’’ the latent variables), so that the vector of predicted responses is given by
\begin{equation} \label{eqn:dec}
\mathbf{r}_i = f_{\text{dec}}(\mathbf{z}_i)
\end{equation}
The two challenges in fitting nonlinear latent variable models are (1) defining and fitting the nonlinear function $f_{\text{dec}}$; and (2) inferring the latent variables $\mathbf{z}_i$, which will depend on the function $f_{\text{dec}}$. We tackle both of these challenges with the use of neural networks (Fig. \ref{fig:srlvm}A). The arbitrary nonlinear function $f_{\text{dec}}$ in equation \eqref{eqn:dec} is modeled with a multi-layer neural network, which can in principle approximate any high-dimensional nonlinear function \cite{hornik1991approximation}. To produce the latent variables $\mathbf{z}_i$, we transform the vector of observed spike counts $\mathbf{y}_i$ into a low-dimensional representation using another arbitrary nonlinear function $f_{\text{enc}}$ (which ``encodes’’ the neural activity into the latent variables), also modeled with a multi-layer neural network:
\begin{equation} \label{eqn:enc}
\mathbf{z}_i = f_{\text{enc}}(\mathbf{y}_i)
\end{equation}
The parameters of the model - the weights and biases of each neural network layer in $f_{\text{enc}}$ and $f_{\text{dec}}$ - can then be learned from experimental data by maximizing the log-likelihood of the model predictions $\mathbf{r}_i$ under an appropriate noise distribution (see Methods), and the value of the latent variables on each trial can be computed using Equation \eqref{eqn:enc}. The resulting model is also referred to as an ``autoencoder'' neural network in the machine learning literature \cite{goodfellow2016deep}, since the full network attempts to reconstruct its inputs.

If the latent variable model defined in Equations \eqref{eqn:dec} and \eqref{eqn:enc} is restricted so that $f_{\text{enc}}$ and $f_{\text{dec}}$ are simple affine transformations (a linear transformation with an additional bias term), and the $\mathbf{z}_i$ are constrained to be nonnegative through the use of pointwise nonlinearities in the hidden layer, the resulting model is the Rectified Latent Variable Model (RLVM) of Whiteway \& Butts \cite{whiteway2017revealing}. We then denote the more general, nonlinear latent variable model, where $f_{\text{enc}}$ and $f_{\text{dec}}$ are multi-layer neural networks, as the ``Stacked’’ RLVM, or SRLVM.

One benefit of the added complexity of the SRLVM is that it can recover nonlinear, low-dimensional structure more parsimoniously than the RLVM. To illustrate this point, we generated the activity of a population of 50 neurons using a nonlinear combination of just four latent variables (see Methods), and fit both models to the data (Fig. \ref{fig:srlvm}B). The performance of the RLVM increases as more latent variables are added, even past four (the true dimensionality of the data). This behavior arises because the inferred latent variables of the RLVM must account not only for the activity of the true latent variables, but also must use additional linear terms to model the nonlinear interactions between them. The SRLVM, on the other hand, is able to model those nonlinear interactions with a neural network. As a consequence, it is able to explain more of the population variability than the RLVM without the additional latent variables.

\subsection{Nonlinear models better describe V1 population response variability}

\begin{figure}[tp]
  \vspace{20pt}
  \centering
  \centerline{\includegraphics[scale=0.9]{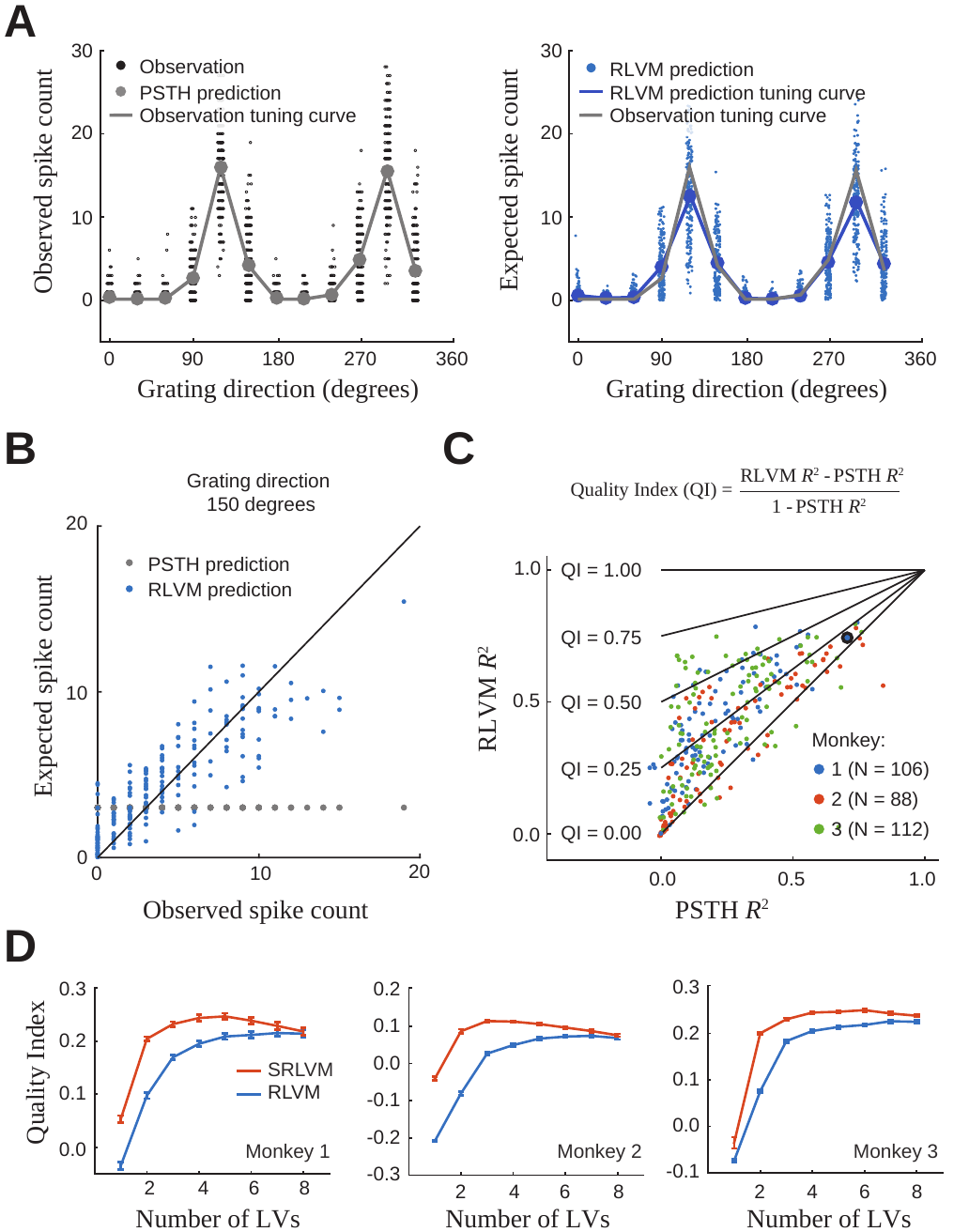}}
  \vspace{5pt}
  \caption{\linespread{1}\selectfont{} \textbf{Population responses from anesthetized macaque V1 are nonlinear and low-dimensional.} \textbf{A}: An illustration of the trial-to-trial variability in an example neuron. \textit{Left}: The observed spike counts (\textit{black dots}) versus grating direction, along with the expected spike counts from the PSTH (\textit{gray dots}). The PSTH predictions are not able to capture the large amount of trial-to-trial variability. \textit{Right}: The expected spike counts from the RLVM (with four latent variables; \textit{blue dots}), which display much more variability. Points have been horizontally jittered in both plots for clarity. \textbf{B}. The expected spike count versus observed spike count for a single grating direction, demonstrating that the trial-to-trial variability in the observations is well captured by the RLVM. \textbf{C}: The Quality Index (QI) compares the predictive ability of a given model (here, the RLVM) to that of the mean stimulus-driven response (PSTH). A high QI results when the model captures a much larger proportion of the variance than the PSTH. A low QI can either result from a poor model, or when the model explains little more than the PSTH. The RLVM $R^2$ versus the PSTH $R^2$ demonstrates the ability of the RLVM to capture trial-to-trial variability across all three datasets. Note that many neurons whose activity are poorly captured by the PSTH can be well predicted by the RLVM. The neuron used in \textbf{A}, \textbf{B} is outlined in black. \textbf{D}: Mean QI over neurons and cross-validation folds, plotted as a function of the number of latent variables in the RLVM (\textit{blue line}) or SRLVM (\textit{red line}). All datasets are best described by a relatively small number of latent variables. \label{fig:v1-srlvm}}
\end{figure}

In order to compare the ability of the RLVM and SRLVM to predict neural responses in experimental data, we fit these models to large population recordings made publicly available by the Kohn Lab and the CRCNS database \cite{kohn2016utah}. This dataset contains Utah array recordings from primary visual cortex (V1) of three anesthetized macaques (Monkeys 1-3 in Figs. \ref{fig:v1-srlvm}, \ref{fig:v1-affine}, \ref{fig:v1-gam}) in response to drifting gratings in 12 equally spaced directions (see Methods for details). We chose this dataset because previous work studying trial-to-trial variability in V1 has shown that a significant amount of this variability is shared among neurons, which has been observed in the context of noise correlations between pairs of neurons \cite{cohen2011measuring}, and using latent variable approaches \cite{ecker2014state, goris2014partitioning, okun2015diverse, lin2015nature, arandia2016multiplicative}.

The first question that we asked is how many linear latent variables are necessary to describe the responses across the neural population. Of course, some of the shared activity is due to the stimulus (i.e. explainable by each neuron's tuning curve) and is identical on repeated trials. Because the RLVM does not use stimulus- or trial-specific information, the response to stimuli will be represented in the latent variables, which may also capture features of the trial-to-trial variability (Fig. \ref{fig:v1-srlvm}A, B). To determine how much of the trial-to-trial variability our models explain, instead of using $R^2$ as a measure for goodness of fit we adopt the Quality Index (QI) introduced in Lin et al. (2015). The QI is a scaled version of $R^2$ (see Methods), such that QI = 0 for models that describe as much variability as the tuning curve, and QI = 1 for perfect prediction (Fig. \ref{fig:v1-srlvm}C). We found that the RLVM was able to explain a substantial portion of the trial-to-trial variability using a small number of latent variables (Fig. \ref{fig:v1-srlvm}D, \textit{blue curves}). These results are similar to what we would find with PCA, because the two models share the same cost function \cite{whiteway2017revealing}. Although the RLVM also incorporates regularization terms which can in principle help with overfitting, we did not find this to be much of a problem given the large number of trials in these datasets (2,400 per session).

As with the simulated data (Fig. \ref{fig:srlvm}), a more compact description of population responses should be possible with a nonlinear latent variable method, given that the data contain nonlinear interactions. To demonstrate this, we fit the SRLVM to the population responses, and indeed the cross-validated performance of the SRLVM is better than that of the RLVM for a given number of latent variables (Fig. \ref{fig:srlvm}D). Similar to the simulation (Fig. \ref{fig:srlvm}B), the nonlinear latent variables not only offer a more compact description of population response variability, but are able to explain more of that variability as well.

Despite the compact representation achieved with the SRLVM, a main limitation in this context is that the model is agnostic to the stimulus tuning of individual neurons, and as a result, the inferred nonlinear latent variables combine features of the stimulus-driven response with shared variability that is not stimulus-driven. This mixing leads to issues of interpretability if we are interested in studying the relationship between stimulus tuning and sources of shared variability. The SRLVM is potentially even more difficult to interpret because it then uses a neural network to nonlinearly transform those mixed latent variables into the predicted responses. As a result, even though the SRLVM can better capture the statistical structure of the data, the resulting model yields latent variables that are difficult to relate to experimentally observable or controllable variables. Nevertheless, the superior performance of the SRLVM suggests that there are indeed nonlinearities present in the data, and as a consequence this model can be used to inform hypotheses about the nature of those nonlinearities. 

\subsection{A neural network-based affine model}

One way to improve interpretability of the models is to posit an explicit form of nonlinear interaction between the stimulus-driven response and the latent variables. One such example of an explicit nonlinearity is multiplication, which previous models have used to capture gain-like modulation of responses to simple stimuli in V1 \cite{goris2014partitioning, lin2015nature, arandia2016multiplicative}. This model, which we will refer to as the ``multiplicative model’’, describes the predicted activity of neuron $n$ on trial $i$, $r_i^n$, in response to a drifting grating at angle $\theta$, as
\begin{equation} \label{eqn:mult}
r_i^n = (1 + w_n g_i) f_n(\theta)
\end{equation}
where $f_n(\theta)$ is the average stimulus-driven response for neuron $n$, $g_i$ is the multiplicative gain term that changes from trial to trial (which is shared across the entire population), and $w_n$ is the coupling weight of neuron $n$ to the gain term. When combined with an additive term, the resulting model describes the predicted response as
\begin{equation} \label{eqn:affine}
r_i^n = (1 + w_n g_i) f_n(\theta) + v_n h_i
\end{equation}
where $h_i$ is an additive offset that, like $g_i$, changes from trial to trial and is shared across the entire population, and $v_n$ is the coupling weight of neuron $n$ to this term. The existence of nonlinearities makes inference of the latent variables much more challenging, and in order to fit these models some of the previous work incorporated particular constraints. For example, Goris et al. (2014) introduced a model similar to that in Equation \eqref{eqn:mult}, with a uniform coupling weight ($w_n = 1$) across all neurons; Lin et al. (2015) introduced the model in Equation \eqref{eqn:affine} as the ``affine’’ model, and again enforced a uniform coupling to the gain term ($w_n = 1$); finally, Arandia-Romero et al. (2016) introduced the ``multi-gain’’ model, where the $w_n$’s and $v_n$’s are unconstrained, but the multiplicative and additive latent variables are constrained to be equal ($g_i = h_i$). 

\begin{figure}[t]
  \vspace{20pt}
  \centering
  \centerline{\includegraphics[scale=0.7]{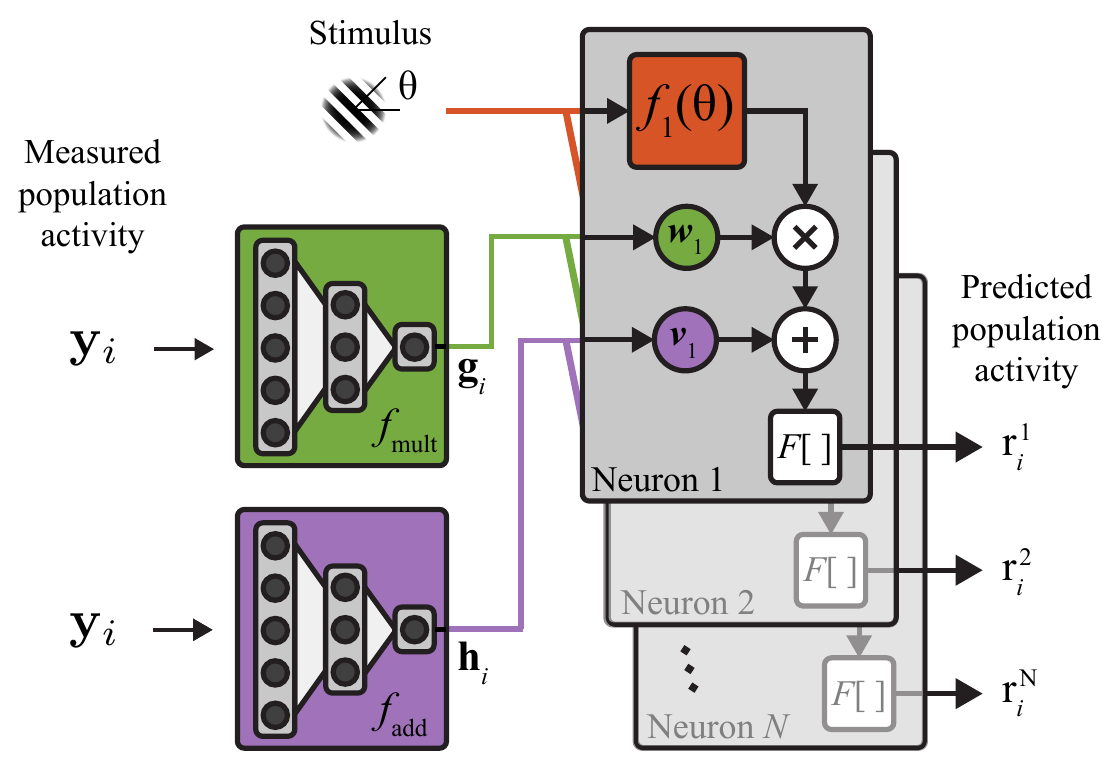}}
  \vspace{5pt}
  \caption{\linespread{1}\selectfont{} \textbf{The structure of the Generalized Affine Model (GAM).} The GAM is a nonlinear latent variable model that explicitly models multiplicative and additive interactions between latent variables and the stimulus-driven response of each neuron. A neural network is used to transform the measured population activity $\mathbf{y}_i$ into multiplicative gain signals $\mathbf{g}_i$ (\textit{green network}). These signals are shared across the entire population and modulate the output of the stimulus-driven response for each individual neuron (\textit{red box}) according to a set of coupling weights $w_n$ for each neuron $n$. A separate neural network transforms the measured population activity into additive signals $\mathbf{h}_i$ (\textit{purple network}), which are also shared across the population, and added to the modulated stimulus-driven response according to a set of coupling weights $v_n$ for each neuron $n$. $F[]$ is an optional spiking nonlinearity that can be applied at the output to arrive at the predicted population activity $\mathbf{r}_i$.}\label{fig:gam}
\end{figure}

Here we can use the framework for inferring latent variables introduced in the previous section to fit the affine model in Equation \eqref{eqn:affine}, imposing no constraints on either the coupling terms or the latent variables. [We will henceforth refer to the model of Lin et al. as the ``constrained’’ affine model.] Our method uses neural networks to transform the observed spike count vector $\mathbf{y}_i$ into each of the latent variables $g_i$ and $h_i$, so that
\begin{eqnarray} 
g_i &=& f_{\text{mult}}(\mathbf{y}_i) \label{eqn:lvs-gam-mult} \\
h_i &=& f_{\text{add}}(\mathbf{y}_i) \label{eqn:lvs-gam-add}
\end{eqnarray}
where $f_{\text{mult}}$ and $f_{\text{add}}$ represent (in general) different neural networks, each with an arbitrary number of layers (Fig. \ref{fig:gam}). As with the SRLVM, all parameters – the weights and biases of the neural networks, and the coupling weights $w_n$, $v_n$, for all $n$ - can be fit simultaneously by maximizing the log likelihood of the model (see Methods), and the value of the latent variables can be computed using Equations \eqref{eqn:lvs-gam-mult} and \eqref{eqn:lvs-gam-add}.

With the ability to fit the fully unconstrained affine model, we first asked how this model compared to some of the constrained versions introduced previously in the literature. One possible outcome is that the constrained model performs equivalently to, or better than, the unconstrained model (potentially due to overfitting of the unconstrained model), a result that supports the validity of the constraints. The other possibility, then, is that the constrained model performs worse than the unconstrained one, in which case the constraints limit the model’s ability to capture the full structure of the data. 

\begin{figure}[t]
  \vspace{20pt}
  \centering
  \centerline{\includegraphics[scale=0.9]{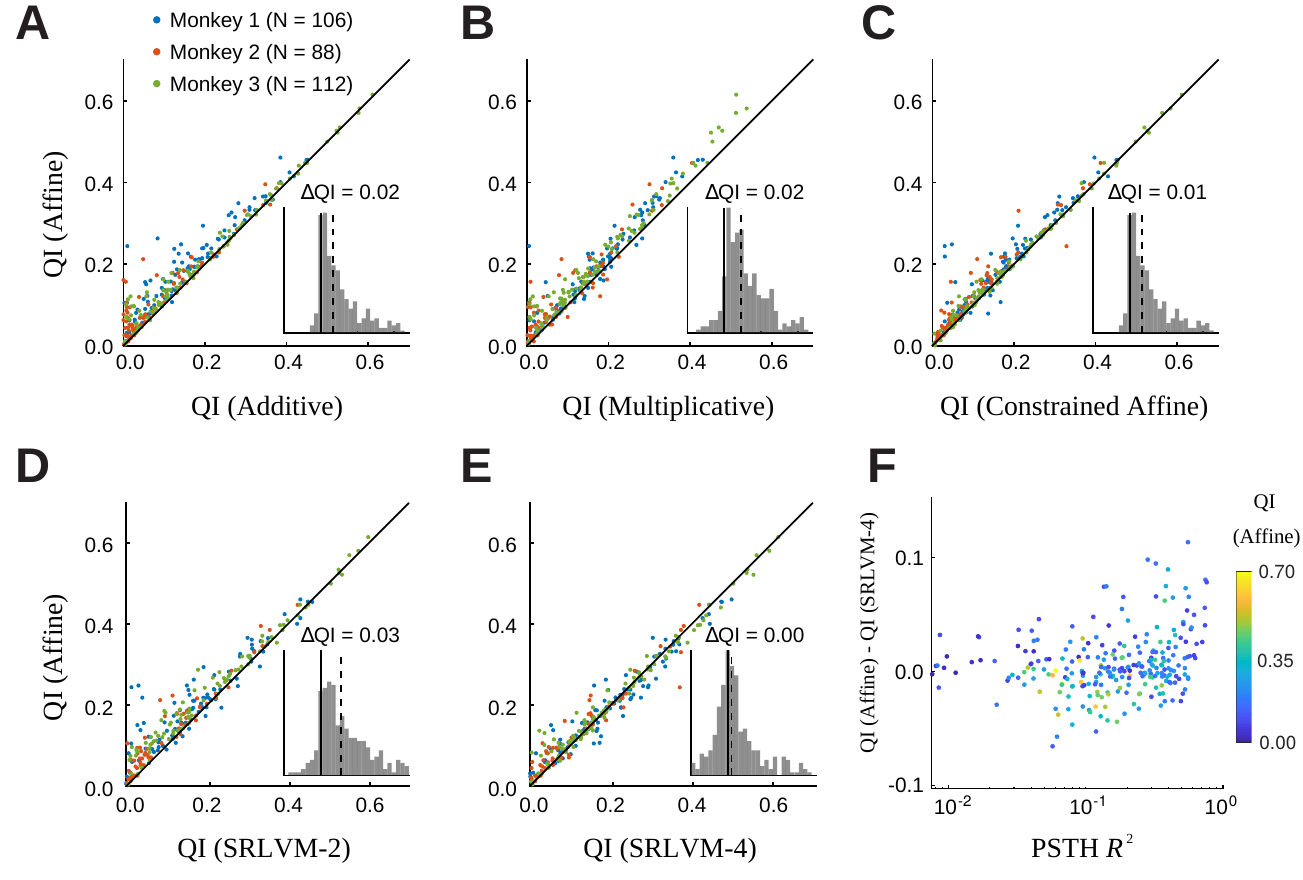}}
  \vspace{5pt}
  \caption{\linespread{1}\selectfont{} \textbf{Performance of additive, multiplicative and affine models in V1 data.} A comparison of the cross-validated performance of the models on individual neurons across three datasets (dataset indicated by color). Each point is the mean QI over 10 cross-validation folds. \textbf{A}: Comparison of the affine model, with one multiplicative and one additive latent variable, to the strictly additive model. \textit{Inset}: A histogram of the difference between the QI of the affine model and the QI of the additive model ($\Delta$QI). Zero is indicated with the solid black line, median difference across all neurons is indicated with the dashed black line. \textbf{B}: Comparison of the affine model to the strictly multiplicative model. \textbf{C}: Comparison of the affine model to the constrained affine model of Lin et al. (2015), which requires all neurons to have the same coupling weight to the multiplicative latent variable. \textbf{D}(\textbf{E}): Comparison of the affine model to the SRLVM with two (four) latent variables. \textbf{F}: The largest increases in affine model QI over the SRLVM-4 QI (y-axis) are for neurons with both large PSTH $R^2$ (x-axis) values and relatively low QI values (color axis), indicating that the affine model is able to better capture the stimulus-driven response with the PSTH model than the SRLVM is able to with latent variables.}\label{fig:v1-affine}
\end{figure}

We found that the unconstrained affine model outperformed both the additive (Fig. \ref{fig:v1-affine}A, p$<$5e-10 for each of the three monkeys, two-sided sign test) and multiplicative models (Fig. \ref{fig:v1-affine}B, p$<$5e-10 for each of the three monkeys). Furthermore, the unconstrained affine model outperformed the constrained affine model (Fig. \ref{fig:v1-affine}C, p$<$5e-5 for each of the three monkeys). Because the effect was small (Fig. \ref{fig:v1-affine}C, \textit{inset}), it suggests that the assumption of the constrained affine model that all neurons have the same coupling to the gain term is a good approximation.

We next asked how well the unconstrained affine model compared to the SRLVM. We first looked at the SRLVM with two latent variables (SRLVM-2), the same number used by the unconstrained affine model. We found that the unconstrained affine model outperformed the SRLVM-2 (Fig. \ref{fig:v1-affine}D, p$<$5e-10 for each of the three monkeys, two-sided sign test). This result is perhaps unsurprising, given that the SRLVM latent variables must capture both the stimulus-driven response and the trial-to-trial variability of the neural population. Increasing the number of SRLVM latent variables to four (after which the SRLVM begins to overfit; see Fig. \ref{fig:v1-srlvm}D) resulted in no significant difference between the unconstrained affine model and the SRLVM (Fig. \ref{fig:v1-affine}E, p$>$0.05 for each of the three monkeys, two-sided sign test).

Although no significant difference exists at the population level, there is still evidence of the difference between these two models at the level of individual neurons. The neurons for which the affine model outperforms the SRLVM-4 are generally well-described by the PSTH and have low affine model QI (Fig. \ref{fig:v1-affine}F). The strong stimulus-driven response of these neurons, then, is not well modeled by the four latent variables of the SRLVM-4, but is adequately captured by the stimulus-driven response term in the affine model. This example demonstrates how models incorporating an explicit form of nonlinear interaction can be compared to more general models such as the SRLVM to understand the extent to which the hypothesized interaction (in this case, multiplication of a latent signal with a stimulus-driven response) is able to capture the statistical structure of the neural responses.

\subsection{Prefrontal cortex is best described by many additive and multiplicative latent variables}

We were surprised that the general nonlinear latent variable models exemplified by the SRLVM did not find more sources of shared variability in V1. Thus, to both test the ability of the nonlinear latent variable models to find more general nonlinear latent variables, and to demonstrate that an affine model should not work in general, we used these models to analyze a dataset from macaque prefrontal cortex \cite{kiani2014dynamics, kiani2015natural}, made publicly available by the Kiani lab at \url{http://www.cns.nyu.edu/kianilab/Datasets.html}. This dataset contains Utah array recordings from three macaques performing a perceptual discrimination task with a random dot motion stimulus of fixed duration (see Methods for details).

\begin{figure}[tp]
  \vspace{20pt}
  \centering
  \centerline{\includegraphics[scale=0.9]{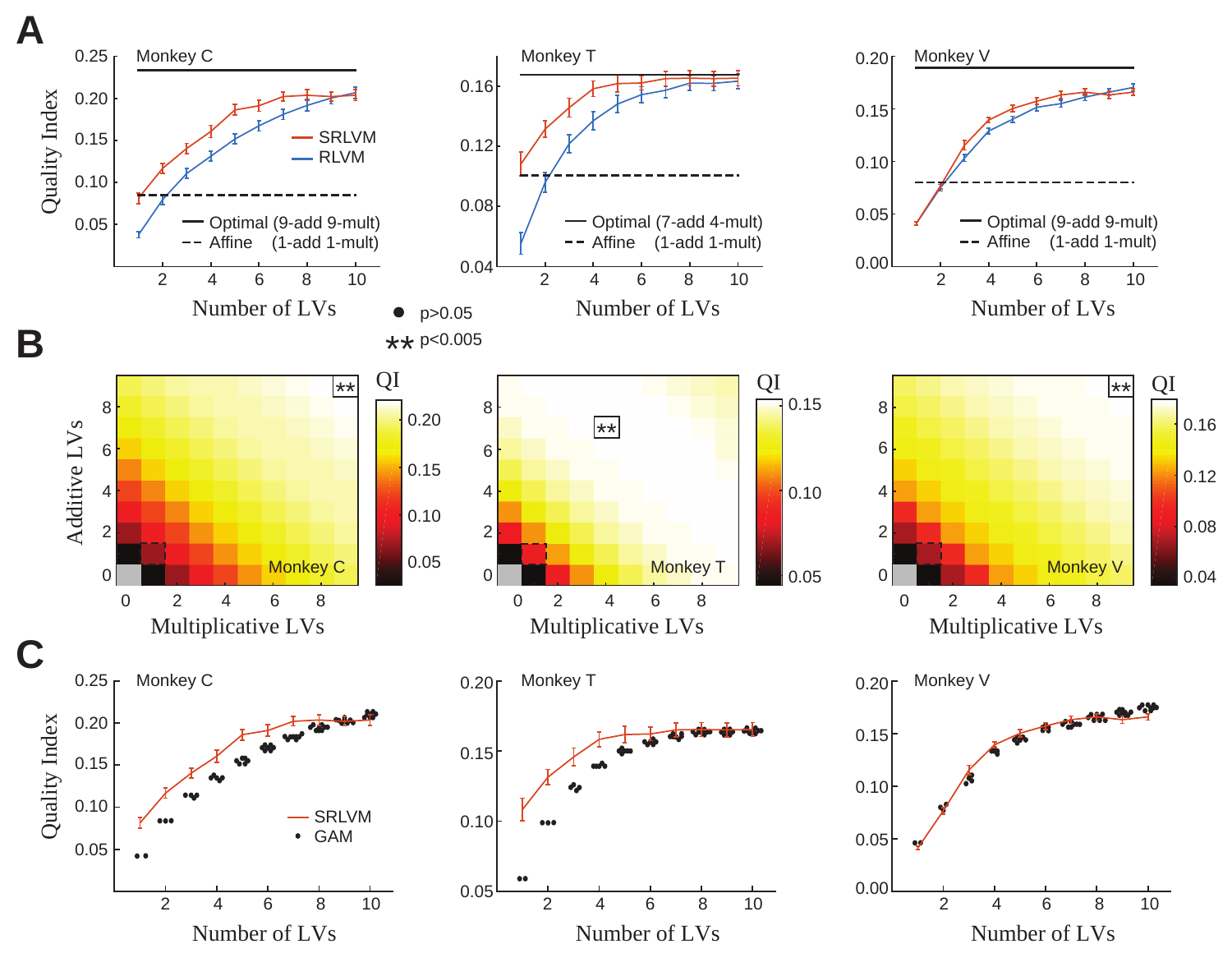}}
  \vspace{5pt}
  \caption{\linespread{1}\selectfont{} \textbf{Population responses from awake PFC data are nonlinear and high-dimensional.} \textbf{A}: Performance of the affine model (\textit{dashed black line}) and the optimal GAM (\textit{solid black line}; see panel \textbf{B}), plotted against the RLVM and SRLVM. The optimal GAM performs approximately as well as the SRLVM, but requires many more latent variables to do so. \textbf{B}: Mean QI over neurons and cross-validation folds for the GAM across a range of additive and multiplicative latent variable combinations. The affine model is outlined (\textit{dashed black line}) as well as the GAM with the highest QI (``optimal GAM’’; \textit{solid black line}; $^{**}p<0.005$, $\bullet>0.05$, two-sided sign test on the mean compared to the affine model). A large number of additive and multiplicative latent variables can describe much more variability than the affine model, in contrast to the V1 data (see Fig. \ref{fig:v1-gam}). \textbf{C}: Comparison of the SRLVM and GAM for a given number of latent variables; for example, a GAM with three latent variables could have 3/0, 2/1, 1/2 or 0/3 additive/multiplicative latent variables. Especially for small numbers of latent variables the SRLVM describes population activity better than the GAM, regardless of the latent variable combinations. GAM points are jittered to display overlapping data points, and error bars have been omitted for clarity. The SRLVM curves are the same as those in panel \textbf{A}.} \label{fig:pfc-gam}
\end{figure}

We found that both the RLVM and SRLVM were able to explain significantly more variability than the affine model (Fig. \ref{fig:pfc-gam}A), in stark contrast to the results from anesthetized V1. First, the best-performing SRLVMs of the PFC population responses have many more latent variables. Furthermore, restricting the SRLVM to two latent variables, in order to match the number in the affine model, still outperforms the affine model in two of the three monkeys, suggesting a more complex nonlinear interaction than the affine model is capable of modeling.

However, the insufficiency of the single additive and multiplicative latent variables of the affine model does not necessarily invalidate the approach of restricting the mathematical form of the latent variable interactions. We extended the affine model to allow an arbitrary number of additive and multiplicative latent variables, which we call the Generalized Affine Model (GAM):
\begin{equation} \label{eqn:gam}
r_i^n = \Bigg(\sum_{k=1}^K w_n^k g_i^k + b_n\Bigg)f_n(\theta) + \sum_{m=1}^M v_n^m h_i^m
\end{equation}
Each neuron is allowed to have its own coupling to each of the $M$ additive latent variables, and a different coupling to each of the $K$ multiplicative latent variables; the unconstrained model is recovered from Equation \eqref{eqn:gam} when $M = K = 1$. Model fitting is performed in exactly the same way as before, by maximizing the log-likelihood of the model predictions.

Using the GAM, we found that a large number of additive and multiplicative latent variables were necessary to describe the PFC population response variability (Fig. \ref{fig:pfc-gam}B). However, the SRLVM was typically able to capture more variance using the same number of latent variables, especially for small numbers of latent variables (comparing, for example, three SRLVM latent variables to one additive and two multiplicative GAM latent variables; Fig. \ref{fig:pfc-gam}C). While the GAM can ultimately match the performance of the SRLVM, the SRLVM can provide a more compact description of activity than the GAM, and thus motivates the search for a different explicit form of nonlinear interactions in this data.

\subsection{V1 remains well-described by the affine model}

\begin{figure}[t]
  \vspace{20pt}
  \centering
  \centerline{\includegraphics[scale=0.9]{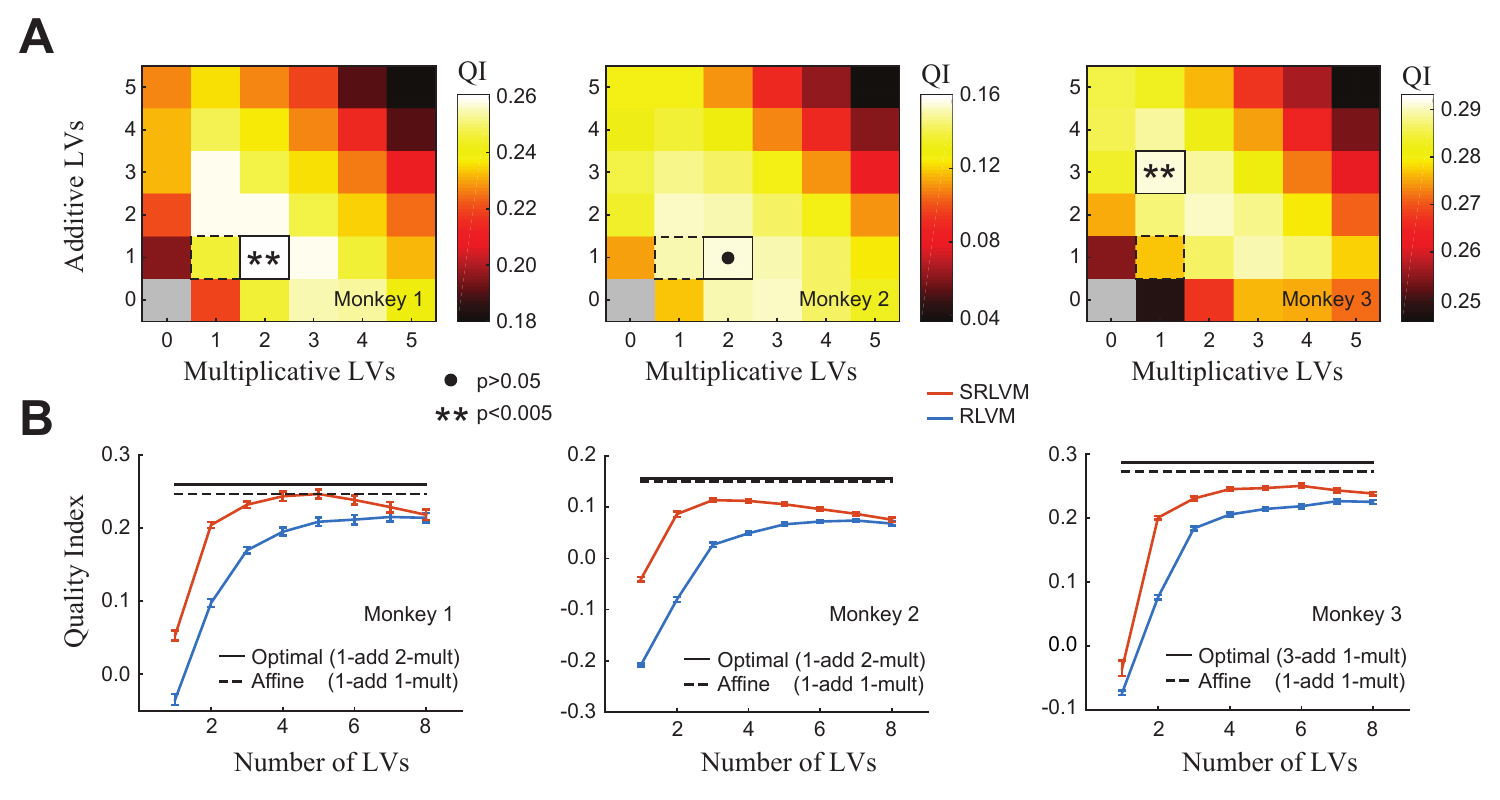}}
  \vspace{5pt}
  \caption{\linespread{1}\selectfont{} \textbf{The affine model is a good description of anesthetized V1 data.} \textbf{A}: Same conventions as Fig. \ref{fig:pfc-gam}B. Only a small number of GAM latent variables are needed to describe the variability, unlike the PFC dataset (compare to Fig. \ref{fig:pfc-gam}B). \textbf{B}: Performance of the affine model (\textit{dashed black line}) and the optimal GAM (\textit{solid black line}), plotted against the RLVM and SRLVM (same curves as Fig. \ref{fig:v1-srlvm}D). The affine model performs almost as well as the optimal GAM, and as well as or better than the SRLVM across all three monkeys, indicating that this model can provide a parsimonious description of the population response.} \label{fig:v1-gam}
\end{figure}

We were also interested in fitting the GAM to the V1 data, where we previously found that the affine model performed comparably to the more flexible SRLVM. Would the GAM be able to explain even more variance? Perhaps unsurprisingly, we found that the best cross-validated models included multiple additive and multiplicative latent variables (Fig. \ref{fig:v1-gam}A). However, the number of additional latent variables in the optimal models was small (1-3 for both types), and the difference was not large in magnitude (Fig. \ref{fig:v1-gam}B, \textit{solid} and \textit{dashed black lines}). This result indicates that the affine model, with a single additive and multiplicative latent variable, is indeed a good description of population response variability in this anesthetized V1 dataset.

\section{Discussion}

We introduced a new framework for fitting nonlinear latent variable models to neural data, based on using neural networks to transform high-dimensional neural activity into low-dimensional latent variables, and proposed two new models within this framework. The first model, the Stacked Rectified Latent Variable Model (SRLVM), is a multi-layer autoencoder neural network that we used as a general nonlinear latent variable model. The SRLVM is agnostic to the stimulus tuning of the neurons, and does not require an explicit form of nonlinear interaction among the latent variables (Fig. \ref{fig:srlvm}A). The second model, the Generalized Affine Model (GAM), extends several previous models of neural population responses  \cite{goris2014partitioning, ecker2014state, lin2015nature, okun2015diverse, arandia2016multiplicative}. The GAM allows for an arbitrary number of additive and multiplicative latent variables, which are shared across the entire population, and modulate the stimulus-driven response of each neuron (Fig. \ref{fig:gam}). For both the SRLVM and the GAM, our model fitting framework allows us to fit all parameters and infer all latent variables simultaneously.

We applied these new models to population recordings from anesthetized macaque V1, and found that the GAM was better able to explain the variance of the population responses with fewer latent variables than the SRLVM (Fig. \ref{fig:v1-gam}B). The optimal GAM explained only slightly more variance than an affine model with a single additive and single multiplicative latent variable (Fig. \ref{fig:v1-gam}B), suggesting that the affine model is a good description of population response variability in this dataset. However, we found the opposite situation in awake macaque PFC, where the SRLVM provided a more compact description of the data than the GAM, and far outperformed the affine model (Fig. \ref{fig:pfc-gam}). These results demonstrate that more general, unconstrained nonlinear latent variable models like the SRLVM can provide a benchmark to test how well more structured nonlinear models (like the affine model) fit the data, and can thus aid in the generation and testing of new hypotheses about the nature of computation in populations of neurons. Furthermore, our comparison between V1 and PFC shows that a range of nonlinear methods will be necessary to study different brain regions under different experimental conditions.

\subsection{Difference between V1 and PFC results}
A main finding presented here was that the PFC neural population requires many more dimensions to describe neural variability, compared with the V1 neural populations. This result could be due to several different factors. First of all, the PFC data was recorded in awake animals performing a behavioral task, unlike the V1 data. Because of this, we cannot make precise statements about how much the difference is due to the awake versus anesthetized state, and how much is due to the different computational requirements of the corresponding tasks. The PFC data was recorded during the execution of a perceptual decision making task, which intuitively places more demands on neural function than passively viewing a stimulus, and might therefore result in higher dimensional neural activity. Gao et al. (2017) \cite{gao2017theory} formalized this intuition and showed that, at least in the linear setting, increased task complexity does indeed result in higher-dimensional neural activity. 

However, it is possible that the differing dimensionalities we observed in these datasets is not related to task complexity at all, and instead related to other non-experiment-related activity present in an awake and behaving animals. Recent work has demonstrated that behavioral variability can drive a large proportion of the variability in rodent cortex, enough even to mask the task-related activity \cite{stringer2018spontaneous, musall2018movement}. It is possible that much of the variability we captured with our models in the PFC data is linked with behavior not directly related to the task, which is absent in the anesthetized V1 data. Fortunately, it is straightforward to incorporate external predictors of activity such as behavior into both the SRLVM and GAM, and so both of these models can be used to study variability while controlling for non-task-related behavior.

\subsection{Advantages and limitations of the modeling framework}
One of the main advantages of our proposed modeling framework is that it is neural network-based. This feature makes the framework highly flexible if one is using software that includes automatic differentiation functionality (such as Tensorflow or PyTorch), since learning model parameters reduces to performing backpropagation. Furthermore, this feature makes the modeling framework more accessible than more specific Bayesian models that may require specialized inference algorithms. It is important to keep in mind, however, that constructing arbitrarily complex neural network mappings can require much more data for training than simpler models. For the SRLVM and GAM we found that the qualitative conclusions made in this work are the same when subsampling neurons and trials (data not shown).

Another advantage that our modeling framework offers (specifically the GAM) is the ability to fit the stimulus-driven responses simultaneously with the latent variables. In all previous work that we have considered here, as well as this work, the datasets contain neural responses to low-dimensional stimuli that could be characterized with simple tuning curves. The GAM expands the range of stimulus sets that can be used since it can incorporate arbitrarily complex stimulus processing models, including Generalized Linear Models \cite{paninski2004maximum} and more complicated, nonlinear models like the Nonlinear Input Model \cite{mcfarland2013inferring} or convolutional neural networks \cite{oliver2016deep, mcintosh2016deep, batty2016multilayer, antolik2016model, cadena2017deep, kindel2017using}. This ability can, for example, allow for the investigation of neural variability during tasks that include more naturalistic stimuli.

\subsection{Assumptions of the modeling framework}
Several other examples of nonlinear latent variable models based on autoencoder neural networks exist in the statistical modeling literature, but differ from our models in the prior assumptions placed on the latent variables. One of the major assumptions of our modeling framework is that the latent variables at time $t$ are a deterministic function of the observed neural activity at time $t$. This assumption has two major consequences: one related to the deterministic mapping, and one related to the modeling of temporal dependencies.

The first consequence is that the latent variables of the SRLVM and GAM are fully deterministic, as in PCA. An alternative is to model the latent variables as random variables distributed according to a specified probability distribution, as in probabilistic PCA (PPCA) and FA. The variational autoencoder (VAE) is a recent advance in statistical modeling \cite{kingma2013auto, rezende2014stochastic} that allows latent variables with relatively simple distributions (e.g. Gaussian) to be mapped through arbitrarily complex neural networks (rather than the simpler affine transformations as in PPCA and FA), resulting in a fully probabilistic, nonlinear latent variable model (where both the latent variables and the observations are described by probability distributions). There are two main advantages to such fully probabilistic models: (1) it is possible to quantify the uncertainty in the inferred latent variables; and (2) the prior distribution placed on the latent variables in such models acts as a regularizer, which can combat overfitting in the small data regime. Although we did not explore the use of fully probabilistic models here (our models specify a probability distribution over the observations but only a point estimate of the latent variables), both the SRLVM and GAM can be cast into a fully probabilistic framework in a manner similar to the VAE, which we leave as a future direction.

The second consequence of the aforementioned assumption (that latent variables at time $t$ are strictly a function of the observed neural data at time $t$) is that there is no notion of dynamics in our models: the data can be randomly permuted in time with no effect on the model fits. Various approaches to fitting dynamical systems to neural data have recently been proposed in the neuroscience literature. The fLDS model developed in \cite{gao2016linear} models the latent variables with a linear dynamical system (LDS) prior, such that the distribution of the latent variables at time $t$ is governed by a linear combination of the latent variables at time $t-1$. LFADS (Latent Factor Analysis via Dynamical Systems) \cite{pandarinath2018inferring} is another model based on the (variational) autoencoder, and uses a more general recurrent neural network to model the dynamics of the latent variables. However, our aim in this work was not to study dynamics (either linear or nonlinear), but to characterize the nonlinear transformations that take place between the latent variables and the observed neural activity. To simplify this endeavor we only modeled spike counts on a per-trial basis, and extending our work to higher time resolution (where dynamics likely play a larger role) is another direction for future investigation.

\subsection{Choice of noise distribution}
The models presented in this work utilize a Gaussian noise distribution over the observations, in contrast to (for example) a Poisson noise distribution. This choice of a noise distribution allowed for explicit comparisons to previous models in the literature \cite{ecker2014state, okun2015diverse, lin2015nature}, and is furthermore reasonable given the large time bins used in the analysis (500-700 ms), combined with the high firing rates of the recorded neurons. Extending the modeling framework to higher time resolutions will require the use of more targeted noise models. 

An interesting question raised by the choice of noise distribution is whether observations of noise in single neurons (e.g., Poisson) will indeed hold as sources of the noise (i.e., latent variables) are modeled explicitly \cite{amarasingham2015ambiguity}. For example, the Poisson distribution fixes the mean and the variance of the observations to be equal, and in many cases spike counts have been found to be over- or under-dispersed (such that the variance is larger than or smaller than the mean, respectively) \cite{goris2014partitioning, stevenson2016flexible, charles2018dethroning}. There have been several approaches to incorporating over- and/or under-dispersion into statistical models of neural activity which are relevant to the models we present here. One approach is to incorporate latent variables that interact with a stimulus-driven response \cite{goris2014partitioning, charles2018dethroning}. For example, the modulated Poisson model of Goris et al. (2014) (referred to here as the multiplicative model) demonstrated that over-dispersion can result from a Poisson noise distribution when a stimulus-independent gain variable multiplies a neuron’s stimulus-driven response (as in the GAM).

Another approach to modeling nonlinear mean-variance relationships has focused on more expressive noise distributions such as the negative binomial \cite{scott2012fully}, generalized count \cite{gao2015high}, and the Conway-Maxwell-Poisson \cite{stevenson2016flexible}. The flexibility of these noise distributions is due to the introduction of additional parameters governing the structure of the distribution; for example, the two-parameter negative binomial distribution is given by $p(y|a, b) \propto
(1 - \sigma(a))^b \sigma(a)^y$, where $\sigma(\cdot)$ is the logistic function. The GAM and SRLVM can naturally incorporate this noise distribution by using the latent variables to parameterize both $a$ and $b$, and again using the log-likelihood as the cost function. The generalized count and Conway-Maxwell-Poisson distributions are not as straightforward to incorporate because they both involve normalization factors that must be computed numerically \cite{stevenson2016flexible}. Nevertheless, it is still possible to use these noise distributions by approximating the normalization factors.

Latent variable models such as the SRLVM and the GAM will be increasingly useful for exploring the growing number of large-scale datasets provided by multi-electrode recordings and two-photon imaging. These models, when used correctly, can capture meaningful structure in the data that can guide our understanding of the principles that underlie brain structure and function, and suggest additional experimental investigations \cite{paninski2017neural}. To fully describe the richness of neural activity, these models must be able to capture nonlinear relationships; to be useful, they must be easily fit to experimental data. The SRLVM and GAM fulfill both of these requirements, and are also easily extendable, allowing them to adapt to changing analysis demands as experimental neuroscience continues to provide datasets of unprecedented size and complexity.

\section{Methods}

\subsection{Experimental data}
\subparagraph{V1 dataset} 
We analyzed electrophysiology data from the Kohn Lab, which has been made publicly available at \url{http://dx.doi.org/10.6080/K0NC5Z4X}. Spiking activity was recorded with a Utah array in primary visual cortex from three anesthetized macaques, in response to full-contrast drifting gratings with 12 equally-spaced directions, presented for 1280 ms (200 repeats). Full details can be found in \cite{smith2008spatial}. Spike counts were analyzed using a single 500 ms time bin per trial (500 ms to 1000 ms after stimulus onset). Spike counts were square-rooted before fitting the models to stabilize the variance and reduce the influence of neurons with high firing rates, as in \cite{byron2009gaussian}.

\subparagraph{PFC dataset} 
We analyzed electrophysiology data from the Kiani Lab, which has been made publicly available at \url{http://www.cns.nyu.edu/kianilab/Datasets.html}. Spiking activity was recorded with a Utah array in area 8Ar of the prearcuate gyrus from three macaques as they performed a direction discrimination task. On each trial the monkey was presented with a random dot motion stimulus for 800 ms, and after a variable-length delay period the monkey reported the perceived direction of motion by saccading to a target in the corresponding direction. The coherence of the dots and their direction of motion varied randomly from trial to trial. Spike counts were analyzed using a single 700 ms time bin per trial during the stimulus presentation (100 ms to 800 ms after stimulus onset). As with the V1 data, spike counts were square-rooted before fitting the models.

\subsection{Simulated data}
We simulated the firing rate $\mathbf{r}_t$ for $N = 50$ neurons from four latent variables, using a two-layer neural network to transform the latent variables into firing rates:
\begin{equation}
\mathbf{r}_t = W_2 f(W_1 \mathbf{z}_t)
\end{equation}
where $f(x)  = \text{max}(0, x)$ is a pointwise nonlinearity. This resulted in population activity that was fully described by a four dimensional space. The four latent variables were each white noise signals, with the value at each time point drawn from a standard normal distribution, and each entry in $W_1$ and $W_2$ was also drawn from a standard normal distribution. The output of the simulation is a continuous-valued firing rate, rather than the resulting spiking activity or two-photon activity, in order to more clearly elucidate model behavior under ideal conditions.

\subsection{Modeling details}
We performed all model fitting using 10-fold cross-validation, where the data are divided into 10 equally-sized blocks, with nine used for training and one for testing, with 10 distinct testing blocks. All reported measures of model performance were calculated using testing data. We used $L_2$ regularization in all of our models to prevent overfitting to the training data, which required fitting a hyperparameter governing the strength of the regularization (e.g. $\lambda$ in Equation \eqref{eqn:srlvm-loss} below). For each fold, we fit the models to the training data using a range of hyperparameter values (six values logarithmically spaced between $10^{-5}$ and $10^0$), and selected the value that resulted in the best performance on the held-out testing fold.

\subparagraph{SRLVM} 
The SRLVM predicts the observed spike count vector of $N$ neurons on trial $i$, $\mathbf{y}­_i \in \mathbb{R}^N$, using a smaller set of latent variables $\mathbf{z}_i \in \mathbb{R}^M$, where typically $M \ll N$. The SRLVM constrains the latent variables to be some encoding function $f_{\text{enc}}$ of the observed spike counts, so that 
\begin{equation}
\mathbf{z}_i = f_{\text{enc}}(\mathbf{y}_i)
\end{equation}
The population activity is then coupled to the latent variables with the decoding function $f_{\text{dec}}$, so that the predicted response vector $\mathbf{r}_i$ of the model is
\begin{equation}
\mathbf{r}_i = f_{\text{dec}}(\mathbf{z}_i)
\end{equation}
We implemented both $f_{\text{enc}}$ and $f_{\text{dec}}$ with feed-forward neural networks, which used rectified linear (ReLU) activation functions $f(x) = \text{max}(0, x)$ as the pointwise nonlinearities in all hidden layers. For all SRLVMs we used three hidden layers, with 10-X-10 number of units, where X corresponds to the number of latent variables. The RLVM has only a single hidden layer corresponding to the latent variables (and thus both $f_{\text{enc}}$ and $f_{\text{dec}}$ are implemented as affine transformations).

To simultaneously fit the weights $\theta_W$ and the biases $\theta_b$ in the networks defining $f_{\text{enc}}$ and $f_{\text{dec}}$, we maximized the log-likelihood of the predicted responses under the Gaussian distribution (with identity covariance matrix), which is equivalent to minimizing the mean square error \cite{bishop2006pattern} between the observed spike counts $\mathbf{y}_i$ and the predicted responses $\mathbf{r}_i$, across all $I$ trials. [Note that it is also possible to use the Poisson log-likelihood as the loss function, which we did not do owing to the large time bins and high spike counts.] We included regularization terms on the network weights to prevent overfitting, so that the final loss function for the model is defined as 
\begin{equation} \label{eqn:srlvm-loss}
\mathcal{L}_{\text{SRLVM}} = \frac{1}{2I}\sum_{i=1}^I \big\| \mathbf{y}_i - \mathbf{r}_i \big\|_2^2 + \lambda q(\theta_W)
\end{equation}
where $q(\cdot)$ is the regularization term that we take to be the $L_2$ norm on the weights, governed by the hyperparameter $\lambda$. Equation \eqref{eqn:srlvm-loss} was optimized using the L-BFGS optimization routine \cite{schmidt2005minfunc}. 

We performed layer-wise initialization of the weights in the encoding network by using varimax-rotated principal components of the observed responses $\{\mathbf{y}_i\}_{i=1}^I$, and initialized weights in the decoding network as transposes of those in the encoding network, which we found to produce models with consistently better cross-validation performance than other initialization schemes (data not shown). 

\subparagraph{GAM} 
The GAM models neural responses using three terms: (1) the stimulus-driven response; (2) one or more gain terms that are shared across the population and multiplicatively modulate the stimulus-driven response; and (3) one or more additive terms that are likewise shared across the population. We define the full model for the predicted response $r_i^n$ of neuron $n$ on trial $i$ in several steps.

We first define a stimulus model $f_n(\mathbf{s}_i)$ for each neuron $n$ such that $f_n$ maps $\mathbf{s}_i$, the vector of stimulus values on trial $i$, to a rate for each $i$. $f_n$ could be as simple as a one-dimensional tuning curve or a peri-stimulus time histogram (PSTH), or any arbitrary feed-forward stimulus processing model like a GLM \cite{paninski2004maximum}, NIM \cite{mcfarland2013inferring} or convolutional neural network \cite{oliver2016deep, mcintosh2016deep, batty2016multilayer, antolik2016model, cadena2017deep, kindel2017using}. In this work, we modeled the stimulus-driven response of each individual neuron in both datasets using an $L_2$-regularized form of the tuning curve, which can be fit using a penalized form of linear regression (ridge regression). We denote the parameters of the stimulus models $\{f_n\}_{n=1}^N$ as $\theta_{\text{stim}}$. For this basic stimulus model, then, the GAM estimates the firing rate $r_i^n$ as
\begin{equation} \label{stimmodel}
r_i^n = F[c_n + f_n(\mathbf{s}_i)]
\end{equation}
where $c_n$ is an overall offset term and $F[\cdot]$ is an optional pointwise spiking nonlinearity (which we take to be linear in our analyses).

Next we define latent variables $\mathbf{g}_i \in \mathbb{R}^K$ that have an explicitly multiplicative effect on the stimulus processing (Fig. \ref{fig:gam} only illustrates a single multiplicative latent variable, but more than one can be fit using this framework). These latent variables are shared across the entire population, though each neuron has its own weight $w_n^k$ to each of the $K$ latent variables $g_i^k$:
\begin{equation} \label{multmodel}
r_i^n = F\Bigg[c_n + u\Bigg(\sum_{k=1}^K w_n^k g_i^k + b_n\Bigg)f_n(\mathbf{s}_i)\Bigg]
\end{equation}
where $b_n$ is a bias term and $u(\cdot)$ is a static nonlinearity. We use $u(x) = 1 + x$ to fit the models in this work, though other functions like $u(x) = \exp(x)$ are also suitable \cite{rabinowitz2015attention}.

Inference of the latent variables is performed by using a neural network $f_{\text{mult}}$ to nonlinearly map the observed population response $\mathbf{y}_i$ into the latent variables:
\begin{equation} \label{gainnn}
\mathbf{g}_i = f_{\text{mult}}(\mathbf{y}_i)
\end{equation}
We denote the parameters for the multiplicative latent variables (${w_n^k}$, ${b_n}$, and the weights and biases of $f_{\text{mult}}$) as $\theta_{\text{mult}}$.

Finally, we allow the GAM to have additive latent variables $\mathbf{h}_i \in \mathbb{R}^M$ as well, in order to capture activity that cannot be accounted for by the modulated stimulus model of equation \ref{gainnn}. Like the multiplicative latent variables, these additive latent variables are shared across the population, but each neuron has its own weight $v_n^m$ to each of the $M$ latent variables $h_i^m$
\begin{equation} \label{eqn:gam-long}
r_i^n = F\Bigg[c_n + u\Bigg(\sum_{k=1}^K w_n^k g_i^k + b_n\Bigg)f_n(\mathbf{s}_i) + \sum_{m=1}^M v_n^m h_i^m\Bigg]
\end{equation}
Inference of the additive latent variables likewise uses a neural network, 
\begin{equation} \label{eq:gam:fadd}
\mathbf{h}_i = f_{\text{add}}(\mathbf{y}_i)
\end{equation}
and we denote the parameters for the additive latent variables (${v_n^m}$ and the weights and biases of $f_{\text{add}}$) as $\theta_{\text{add}}$. 

We implemented both $f_{\text{mult}}$ and $f_{\text{add}}$ with affine transformations; including hidden layers, such that the transformation from observed activity to latent variables is nonlinear, did not result in substantive model improvements (data not shown), likely due to the increased number of parameters and relative lack of data.

To fit the parameters $\{\theta_{\text{stim}}, \theta_{\text{mult}}, \theta_{\text{add}}\}$ of the GAM we define the loss function to be the penalized negative log-likelihood $\mathcal{L}_{\text{GAM}}$ under the Gaussian noise model (as with the SRLVM):
\begin{equation} \label{negloglike}
\mathcal{L}_{\text{GAM}} = \frac{1}{2I}\sum_{i=1}^I \big\| \mathbf{y}_i - \mathbf{r}_i \big\|_2^2 + \lambda_{\text{stim}} q(\theta_{\text{stim}}) + \lambda_{\text{add}} q(\theta_{\text{add}}) + \lambda_{\text{mult}} q(\theta_{\text{mult}})
\end{equation}
where the $q(\cdot)$ are regularization terms that we take to be the $L_2$ norm on the weights, governed by the hyperparameter $\lambda$.  
The parameters $\theta_{\text{mult}}$ and $\theta_{\text{add}}$ are initialized using the varimax-rotated principal components of the observed responses $\{\mathbf{y}_i\}_{i=1}^I$; $\theta_{\text{stim}}$ is initialized by fitting the desired stimulus processing model $f_n$ (equation \ref{stimmodel}) to each neuron individually and using the resulting parameters. To train the full model we hold $\theta_{\text{stim}}$ fixed and simultaneously optimize $\theta_{\text{mult}}$ and $\theta_{\text{add}}$ using the L-BFGS optimization routine \cite{schmidt2005minfunc}.

Note that the coupling weights and latent variables of the GAM are unidentifiable since, for any nonzero scalar $\alpha$, the product $w_n^k g_i^k$ in Equation \eqref{eqn:gam-long} is equivalent to $(w’)_n^k (g’)_i^k$, where $(w’)_n^k = \alpha w_n^k$ and $(g’)_i^k = (1 / \alpha) g_i^k$ (and the same observation holds for the additive latent variables). The $L_2$ regularization that we place on the network weights limits the range of values of the $w_n^k$, but even with this penalty the whole subspace of the inferred latent variables is the relevant object for further investigation, rather than the specific magnitudes of the latent variables or the axes of the subspace.

\subsection{Evaluating model performance}
To quantify the goodness-of-fit of the different models we employed the coefficient of determination across the full population, defined as
\begin{equation}
R^2 = \frac{1}{N}\sum_{n=1}^N \Bigg[1 - \frac{\sum_i(y_i^n - r_i^n)^2}{\sum_i(y_i^n - \bar{y}^n)^2}\Bigg]
\end{equation}
where $\bar{y}^n$ is the average value for neuron $n$ across all trials.

To actually evaluate model performance of the SRLVM we performed a version of the leave-one-out method introduced in \cite{byron2009gaussian}: first the full model is fit to all of the training data, and then for cross-validation the estimated population activity $r_i^n$ on trial $i$ for neuron $n$ is calculated as
\begin{equation}
r_i^n = [f_{\text{dec}}(f_{\text{enc}}(\mathbf{y}_i^{-n}))]_n
\end{equation}
where $\mathbf{y}_i^{-n}$ is the population activity on trial $i$ with the value for neuron $n$ set to zero. The prediction $r_i^n$ is calculated in this way for each value of $i$ and for each neuron $n$, and all such values are then combined for the final prediction $\{\mathbf{r}_i\}_{i=1}^I$. This procedure results in low cross-validation performance if any single neuron dominates the activity of a latent variable. The same procedure is used for GAMs, so that
\begin{equation}
r_i^n = F\Bigg[c_n + u\Bigg(\mathbf{w}_n^{T} f_{\text{mult}}(\mathbf{y}_i^{-n}) + b_n\Bigg)f_n(\mathbf{s}_i) + \mathbf{v}_n^T f_{\text{add}}(\mathbf{y}_i^{-n}) \Bigg]
\end{equation}
where $\mathbf{w}_n$ and $\mathbf{v}_n$ are the multiplicative and additive coupling terms for neuron $n$ collected into vectors.

To compare the performance of the SRLVM and GAM, we compare the $R^2$ of the given latent variable model (``MODEL'') with the $R^2$ of the stimulus model (``STIM''):
\begin{equation}
\text{QI} = \frac{R_{\text{MODEL}}^2 - R_{\text{STIM}}^2}{1 - R_{\text{STIM}}^2}
\end{equation}
This measure will be equal to zero when the latent variable model explains as much variability as the stimulus model, and one when the model predicts the neural activity perfectly.

\section*{acknowledgements}
We thank both the Kohn and Kiani labs for making their data publicly available, and CRCNS for hosting the Kohn dataset.

\section*{conflict of interest}
No conflicts of interest, financial or otherwise, are declared by the authors.

\section*{Online materials}
All data and code used to produce the figures are available online. The V1 dataset can be accessed through the CRCNS database (\url{http://dx.doi.org/10.6080/K0NC5Z4X}) and the PFC dataset can be accessed through the Kiani Lab website (\url{http://www.cns.nyu.edu/kianilab/Datasets.html}). The code is available at \url{https://github.com/themattinthehatt/whiteway-et-al-2019-nbdt}.

\printendnotes

\bibliography{main}

\end{document}